\documentstyle[epsfig]{aipproc}
\newcommand{\be}{\begin{eqnarray}}
\newcommand{\ba}{\begin{array}}
\newcommand{\ea}{\end{array}}
\newcommand{\ee}{\end{eqnarray}}
\newcommand{\dslash}{\partial \hskip -0.5em /}

\begin{document}
\rightline{UNITU-THEP-10/1997}
\title{Structure Functions from Chiral Soliton Models\thanks{Talk
presented by HW at the $6^{\rm th}$ {\it Conf. on Intersections
between Nuclear and Particle Physics}, Big Sky, Mt, 
May $27{\rm th}$--June $2^{\rm nd}$, 1997.  Work supported
in part by the Deutsche Forschungsgemeinschaft (DFG) under 
contract Re 856/2--3 and the US--DOE grant DE--FE--02--95ER40923.}}

\author{H. Weigel$^*$, L. Gamberg$^{\dagger}$,
and H. Reinhardt$^*$}
\vspace{-0.2cm}
\address{$^*$Institute for Theoretical Physics, T\"ubingen University\\
Auf der Morgenstelle 14, D-72076 T\"ubingen, Germany\\
$^{\dagger}$Department of Physics and Astronomy, University of
Oklahoma\\
440 W. Brooks Ave, Norman, Oklahoma 73019--0225, USA}

\maketitle

\vspace{-0.3cm}

\begin{abstract}
We study nucleon structure functions within the bosonized
Nambu--Jona--Lasinio (NJL) model where the nucleon emerges as a
chiral soliton.  We discuss the model predictions on the Gottfried 
sum rule for electron--nucleon scattering. A comparison with a 
low--scale parametrization shows that the model reproduces the gross 
features of the empirical structure functions. We also compute the 
leading twist contributions of the polarized structure functions $g_1$ 
and $g_2$ in this model. We compare the model predictions on these 
structure functions with data from the E143 experiment by GLAP evolving 
them from the scale characteristic for the NJL-model to the scale of 
the data. 
\end{abstract}

The purpose of this investigation is to provide a link between two 
successful although seemingly unrelated pictures of baryons. On one 
side we have the quark parton model which successfully describes 
the scaling behavior of the structure functions in deep inelastic 
scattering (DIS) processes. The deviations from these scaling laws 
are computable in the framework of perturbative QCD. On the other side 
we have the chiral soliton approach which is motivated by the large 
$N_C$ expansion of QCD, $N_C$ being the number of color degrees of 
freedom. For $N_C\to\infty$, QCD is known to be equivalent to an 
effective theory of weakly interacting mesons. Although this theory 
is not explicitly known it can be modeled by assuming that at low 
energies only the light mesons (pions, kaons, $\rho$, $\omega$) are 
relevant. When modeling the meson theory one requires the symmetry 
structure of QCD. In particular besides Pioncar\`e invariance we 
require chiral symmetry and its spontaneous breaking. Baryons  
emerge as non--perturbative (topological) configurations of the 
meson fields, the so--called solitons. The link between these two 
pictures can be established by computing structure functions within 
a chiral soliton model for the nucleon from the hadronic tensor  
\be
\vspace{-0.1cm}
W^{ab}_{\mu\nu}(q)=\frac{1}{4\pi}\int d^4 \xi \
{\rm e}^{iq\cdot\xi}
\langle N(P) |\left[J^a_\mu(\xi),J^{b{\dag}}_\nu(0)\right]|N(P)\rangle \ ,
\label{deften}
\ee
which describes the strong interaction part of the DIS 
cross--section. In eq (\ref{deften}) $|N(P)\rangle$ refers to the 
nucleon state with momentum $P$ and $J^a_\mu(\xi)$ to the hadronic 
current suitable for the process under consideration. In most
soliton models -- due to the non--perturbative nature of the soliton 
configuration -- the current commutator (\ref{deften}) remains 
intractable. However, the Nambu and Jona--Lasinio (NJL) model 
\cite{Na61} of quark flavor dynamics, which can be bosonized by 
functional integral techniques \cite{Eb86}, contains simple current 
operators. Most importantly, the bosonized version of the NJL--model 
contains soliton solutions \cite{Al96}. This paves the way to compute
structure functions in the soliton approach. 

In order to extract the leading twist contributions to the structure 
function one computes the hadronic tensor in the Bjorken limit
\be
\vspace{-0.1cm}
q_0=|\mbox{\boldmath $q$}| - M_N x
\quad {\rm with}\quad
|\mbox{\boldmath $q$}|\rightarrow \infty 
\quad {\rm and}\quad
x={-q^2}/{2P\cdot q}\quad  {\rm fixed}\ .
\label{bjlimit}
\ee
Here we confine ourselves to presenting the key issues of the 
calculation, details may be traced from refs. 
\cite{We96,We97a,We97b}.

\vspace{-0.2cm}
\section*{The Nucleon from the Chiral Soliton in the NJL Model}
\vspace{-0.2cm}

In this section we briefly summarize the basic features of 
the chiral soliton in the NJL--model and discuss how states with 
nucleon quantum numbers are generated. For more details see
refs. \cite{Al96,Ch96} and quotations therein.

The NJL--model Lagrangian contains a quartic quark interaction 
which is chirally symmetric. Derivatives of the quarks fields 
only appear in form of a free Dirac Lagrangian, hence the current 
operator is formally free. Upon bosonization the action may 
be expressed as \cite{Eb86}
\be
\vspace{-0.1cm}
{\cal A}={\rm Tr}{\rm ln}_\Lambda
\left(i\dslash - m U^{\gamma_5}\right)
+\frac{m_0m}{4G}{\rm tr}\left(U+U^{\dag}-2\right)
\label{bosact}
\ee
where we have confined ourselves to the interaction in the 
pseudoscalar channel. The associated pion fields 
$\mbox{\boldmath $\pi$}$ are contained in the non--linear 
realization 
$U={\rm exp}(i\mbox{\boldmath $\tau$}\cdot
\mbox{\boldmath $\pi$}/f_\pi)$. In eq (\ref{bosact}) ${\rm tr}$ 
denotes discrete flavor trace while ${\rm Tr}$ also includes
the functional trace. The parameters of the model are the coupling 
constant $G$, the current quark mass $m_0$ and the UV cut--off 
$\Lambda$. The constituent quark mass $m$ arises as the solution to 
the Schwinger--Dyson (gap) equation and characterizes 
the spontaneous breaking of chiral symmetry. A Bethe--Salpeter 
equation of the pion field can be derived from eq (\ref{bosact}) 
which allows one to express the pion mass $m_\pi=135{\rm MeV}$ and 
decay constant $f_\pi=93{\rm MeV}$ in terms of the model parameters.
Fixing these quantities leaves one parameter undetermined which 
maybe expressed in terms of the constituent quark mass $m$.
Subsequently an energy functional for non--perturbative but
static field configurations $U(\mbox{\boldmath $r$})$ can be 
extracted from (\ref{bosact}). It can be expressed as a 
regularized sum of single quark energies $\epsilon_\mu$. For 
the hedgehog {\it ansatz}, 
$U_H={\rm exp}(i\mbox{\boldmath $\tau$}\cdot
{\hat{\mbox{\boldmath $r$}}}\Theta(r))$ 
the assoicated 
one--particle Dirac Hamiltonian becomes
\be
h=\mbox{\boldmath $\alpha$}\cdot\mbox{\boldmath $p$}
-\beta\ m\ {\rm exp}\left(i\gamma_5\mbox{\boldmath $\tau$}\cdot
{\hat{\mbox{\boldmath $r$}}}\Theta(r)\right)\ , \quad
h\Psi_\mu = \epsilon_\mu\Psi_\mu \ .
\label{dirham}
\ee
The distinct level (v), which is bound in the background of $U_H$,
is referred to as the valence quark state. Its explicit 
occupation guarantees unit baryon number. The chiral angle 
$\Theta(r)$ of the soliton is determined by self--consistently 
minimizing the energy functional. This soliton configuration does 
not yet carry nucleon quantum numbers. To generate them the (unknown) 
time dependent field configuration is approximated by elevating the 
zero modes to time dependent collective coordinates
$U(\mbox{\boldmath $r$},t)=A(t)U_H(\mbox{\boldmath $r$})A^{\dag}(t), 
\ A(t)\in {\rm SU}(2).$
Upon canonical quantization the angular velocities,
$\mbox{\boldmath $\Omega$}=-2i{\rm tr}
(\mbox{\boldmath $\tau$}A^{\dag}\dot A)$, are replaced by the 
spin operator $\mbox{\boldmath $J$}$ via 
$\mbox{\boldmath $\Omega$}=\mbox{\boldmath $J$}/\alpha^2$
with $\alpha^2$ being the moment of inertia\footnote{Generalizing this
treatment to flavor SU(3) indeed shows that the baryons have to be 
quantized as half--integer objects. For a review on solitons in 
SU(3) see {\it e.g.} \cite{We96a}.} while the nucleon states
$|N\rangle$ emerge as Wigner $D$--functions.
To compute nucleon properties the action (\ref{bosact}) is expanded 
in powers of $\mbox{\boldmath $\Omega$}$ corresponding to an
expansion in $1/N_C$. In particular the valence quark wave--function
$\Psi_{\rm v}(\mbox{\boldmath $x$})$ acquires a linear correction 
\be
\Psi_{\rm v}(\mbox{\boldmath $x$},t)=
{\rm e}^{-i\epsilon_{\rm v}t}A(t)
\left\{\Psi_{\rm v}(\mbox{\boldmath $x$})
+\sum_{\mu\ne{\rm v}}
\Psi_\mu(\mbox{\boldmath $x$})\hspace{-2pt}
\frac{\langle \mu |\mbox{\boldmath $\tau$}\cdot
\mbox{\boldmath $\Omega$}|{\rm v}\rangle}
{2(\epsilon_{\rm v}-\epsilon_\mu)}\right\}=
{\rm e}^{-i\epsilon_{\rm v}t}A(t)
\psi_{\rm v}(\mbox{\boldmath $x$}).
\label{valrot}
\ee
Here $\psi_{\rm v}(\mbox{\boldmath $x$})$ refers to the spatial part
of the body--fixed valence quark wave--function with the rotational
corrections included.

\vspace{-0.2cm}
\section*{Structure Functions in the Valence Quark Approximation}
\vspace{-0.2cm}

The starting point for computing the unpolarized structure
functions is the symmetric part of hadronic tensor in a form 
suitable for localized fields \cite{Ja75}, 
\be
W^{lm}_{\{\mu\nu\}}(q)&=&\zeta\int \frac{d^4k}{(2\pi)^4} \
S_{\mu\rho\nu\sigma}\ k^\rho\
{\rm sgn}\left(k_0\right) \ \delta\left(k^2\right)
\int_{-\infty}^{+\infty} dt \ {\rm e}^{i(k_0+q_0)t}
\nonumber \\ && \hspace{0.1cm}
\times \int d^3x_1 \int d^3x_2 \
{\rm exp}\left[-i(\mbox{\boldmath $k$}+\mbox{\boldmath $q$})\cdot
(\mbox{\boldmath $x$}_1-\mbox{\boldmath $x$}_2)\right]
\nonumber \\ && \hspace{0.1cm}
\times \langle N |\left\{
{\hat{\bar \Psi}}(\mbox{\boldmath $x$}_1,t)t_l t_m\gamma^\sigma
{\hat\Psi}(\mbox{\boldmath $x$}_2,0)-
{\hat {\bar \Psi}}(\mbox{\boldmath $x$}_2,0)t_m t_l\gamma^\sigma
{\hat\Psi}(\mbox{\boldmath $x$}_1,t)\right\}| N \rangle .
\label{stpnt}
\ee
Note that the quark spinors are functionals of the soliton.
Here $S_{\mu\rho\nu\sigma}=g_{\mu\rho}g_{\nu\sigma}
+g_{\mu\sigma}g_{\nu\rho}-g_{\mu\nu}g_{\rho\sigma}$ and
$\zeta=1(2)$ for the structure functions associated with the
vector (weak) current and $t_m$ is a suitable isospin 
matrix. The matrix element between the nucleon states 
($|N\rangle$) is taken in the space of the collective coordinates. 
In deriving eq. (\ref{stpnt}) the {\it free} correlation function 
for the intermediate quark fields has been assumed. In the Bjorken 
limit (\ref{bjlimit}) the momentum, $k$, of the intermediate quark 
is highly off--shell and hence not sensitive to momenta typical 
for the soliton configuration. Thus the use of the free correlation 
function is a valid treatment in this kinematical regime. 

\begin{figure}
\centerline{
\hspace{-0.5cm}
\epsfig{figure=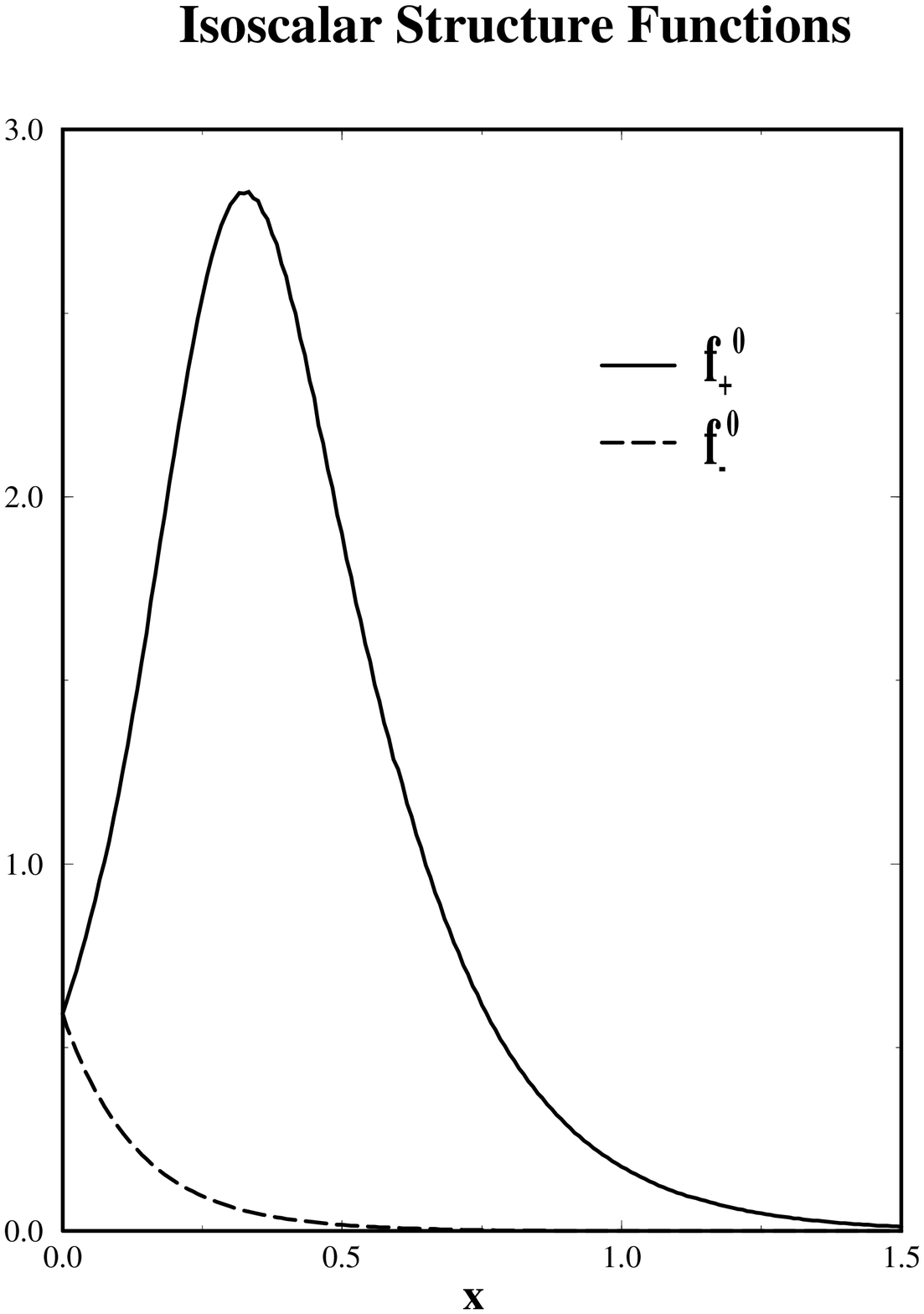,height=5.0cm,width=7.0cm}
\hspace{-1.0cm}
\epsfig{figure=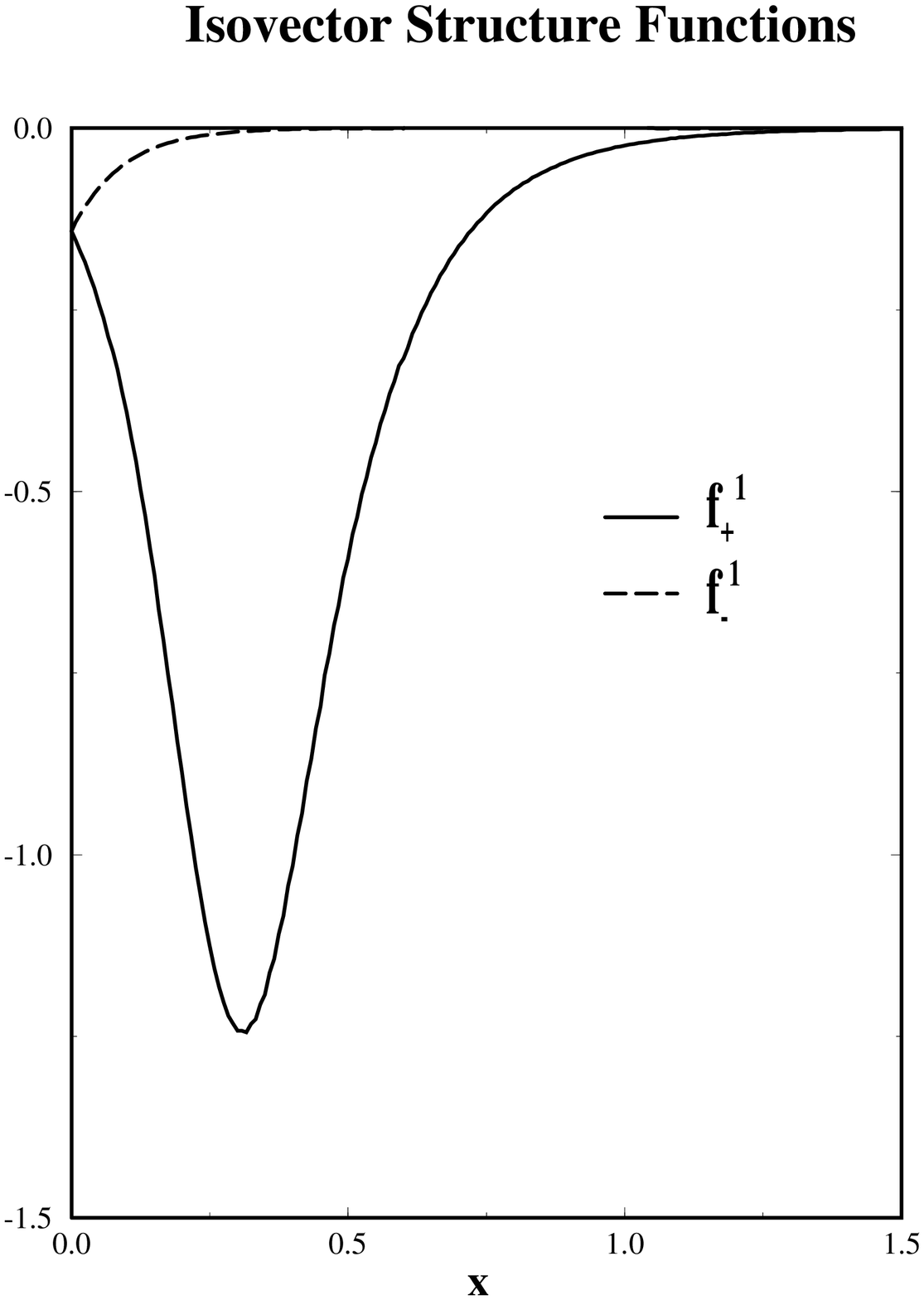,height=5.0cm,width=7.0cm}}
~
\vspace{0.2cm}
\caption{\label{fig_1}The unpolarized structure functions
obtained after extracting the collective part of the
nucleon matrix elements. Here we used $m=350{\rm MeV}$.}
\vspace{-.2cm}
\end{figure}

The valence quark approximation ignores the vacuum polarization in 
(\ref{stpnt}), {\it e.g.} the quark field operator ${\hat\Psi}$ is 
substituted by the valence quark contribution (\ref{valrot}). For 
small constituent quark masses $m\sim400{\rm MeV}$ this is well 
justified since this level provides the dominant share to static 
observables \cite{Al96,Ch96}. The structure function $F_2(x)$ can be 
obtained from (\ref{stpnt}) by an appropriate projection\footnote{In 
the Bjorken limit the Callan--Gross relation $F_2(x)=2x F_1(x)$ is 
satisfied.}. After computing the collective coordinate matrix 
elements all physical relevant processes are described in terms 
of four structure functions $f_\pm^{0,1}$. The superscript denotes 
the isopsin combination of $t_l t_m$ while the subscript refers to
forward and backward moving intermediate quarks in (\ref{stpnt}).
In figure \ref{fig_1} the predictions for these four structure 
functions are displayed. 
Although the problem is not formulated Lorentz--covariantly these 
structure functions are reasonably well localized in the interval 
$x\in [0,1]$. Furthermore the contributions of the backward moving 
quarks are quite small, however, they increase with $m$. Note that 
for consistency with the Adler sum rule also the moment of inertia 
must be restricted to the valence quark contribution 
\cite{We96,We97a}. For $m=350{\rm MeV}$ this, however, is 
almost 90\%.

\begin{figure}
\centerline{
\epsfig{figure=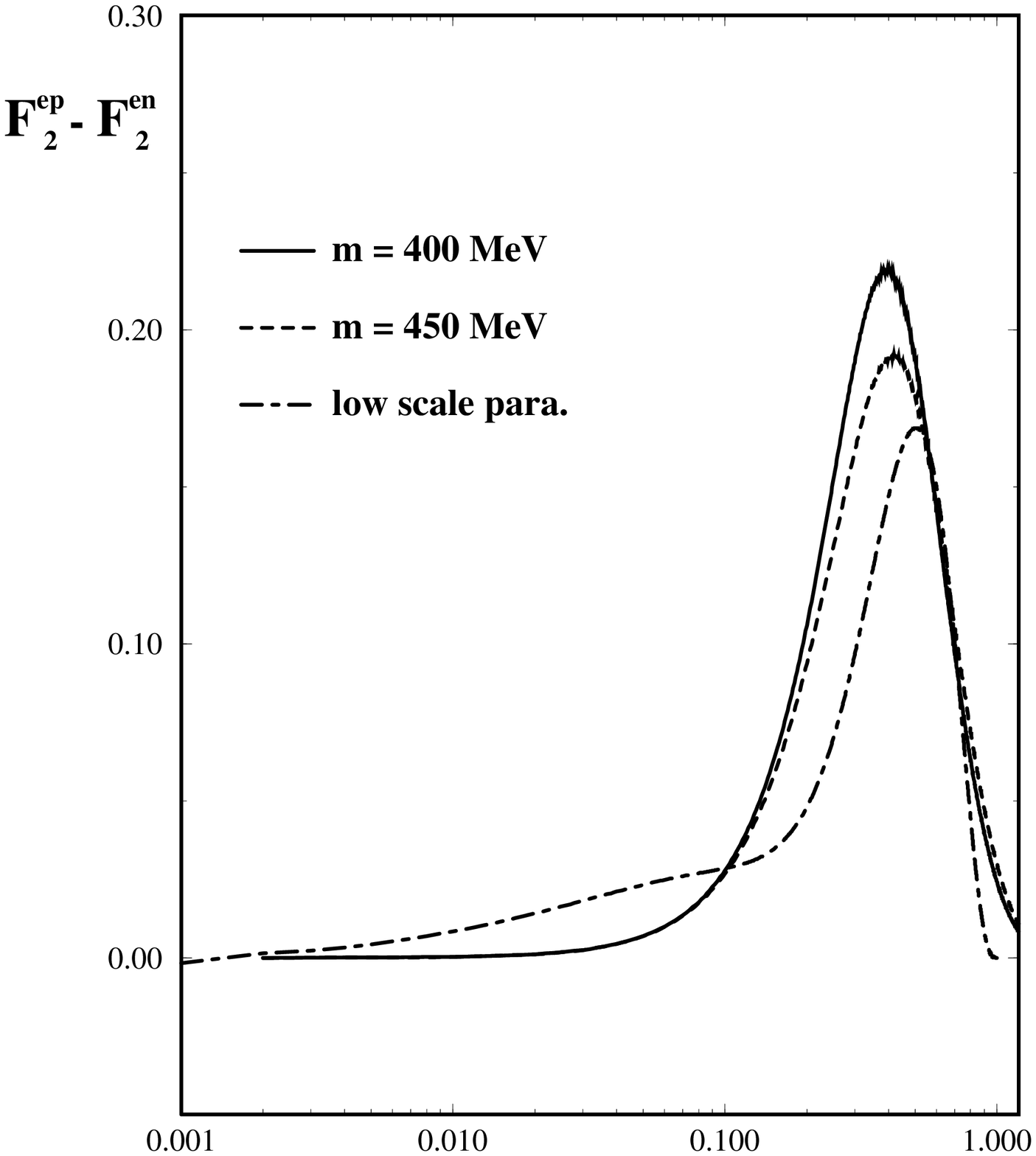,height=5.0cm,width=7.0cm}
\hspace{-0.5cm}
\epsfig{figure=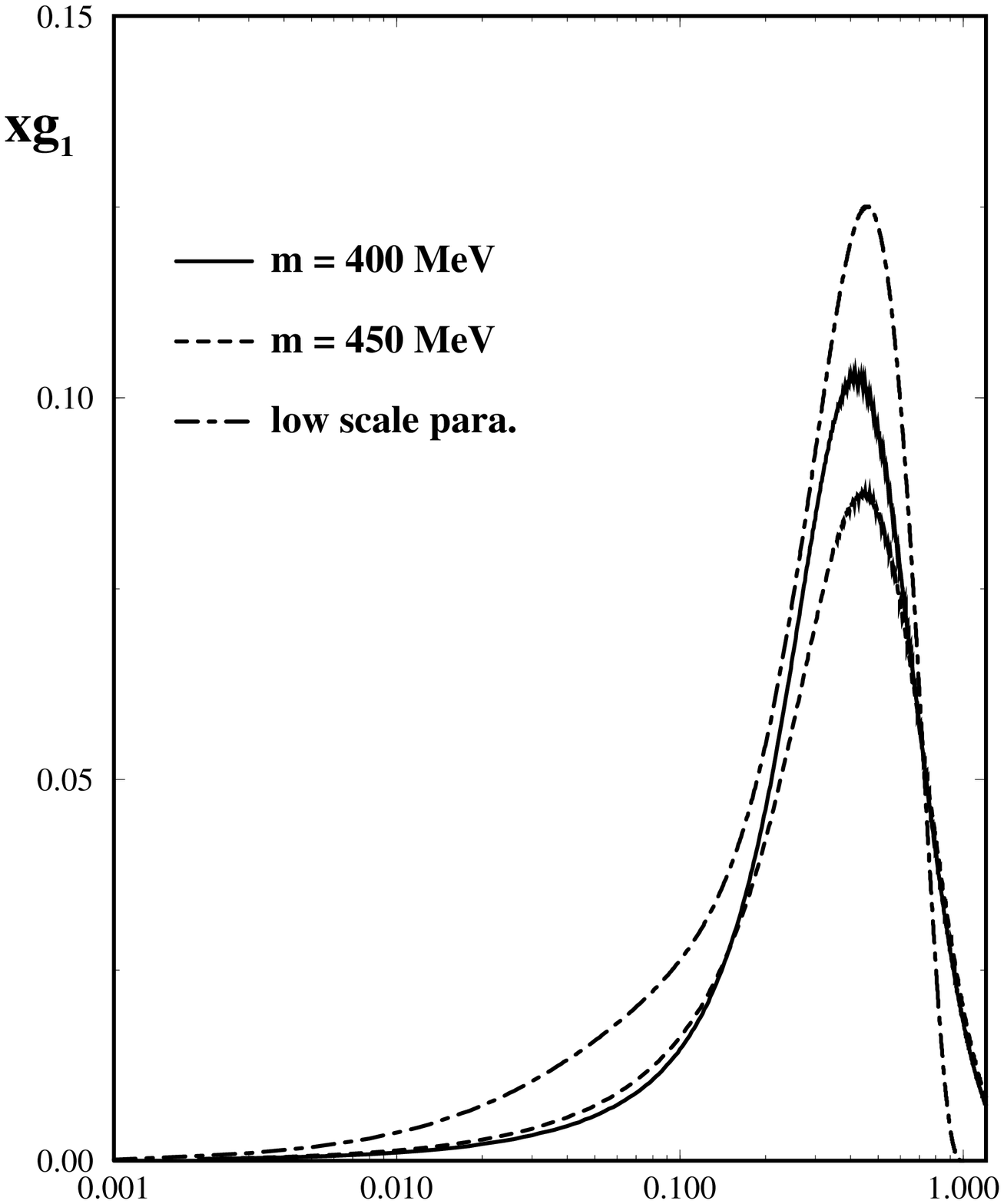,height=5.0cm,width=7.0cm}}
~
\vspace{0.2cm}
\caption{\label{fig_2}Comparison of the model structure functions 
with the low--scale parametrization of ref \protect\cite{Gl95}.
Left panel: The structure function $F_2(x)$ for 
electron--nucleon scattering. Right panel: The polarized structure
function $xg_1$ for the nucleon.} 
\vspace{-.2cm}
\end{figure}

We continue by presenting the numerical results for the 
structure functions for physical processes. In figure \ref{fig_2} 
we display the linear combination relevant for the Gottfried sum rule
\be
\left(F_2^{ep}-F_2^{en}\right)=-
x\left(f_+^1-f_-^1\right)/3
\label{gott}
\ee
and compare it to the low--scale parametrization of the empirical
data \cite{Gl95}. This is obtained from a next--to--leading
order QCD evolution of the experimental to a low--energy regime, where 
soliton models are valid. The agreement improves with
increasing constituent quark masses.
Apparently the model reproduce the gross features of the 
low--scale parametrization. Moreover the integral of the 
Gottfried sum rule
\be
S_G=
\int_0^\infty \frac{dx}{x}
\left(F_2^{ep}-F_2^{en}\right) 
=\cases{0.29\ , \ m=400 {\rm MeV}\cr
0.27\ , \ m=450 {\rm MeV}}
\label{gottrule}
\ee
agrees reasonably well with the empirical value $S_G=0.235\pm0.026$
\cite{Ar94}. In particular the deviation from the na{\"\i}ve value 
(1/3) \cite{Go67} is in the direction demanded by experiment. 

Figure \ref{fig_2} also shows the comparison of the model prediction for 
the polarized structure function $g_1(x)$ with the corresponding 
low--scale parametrization \cite{Gl95}.
In this case the agreement improves with decreasing $m$. 

\begin{figure}
\centerline{
\epsfig{figure=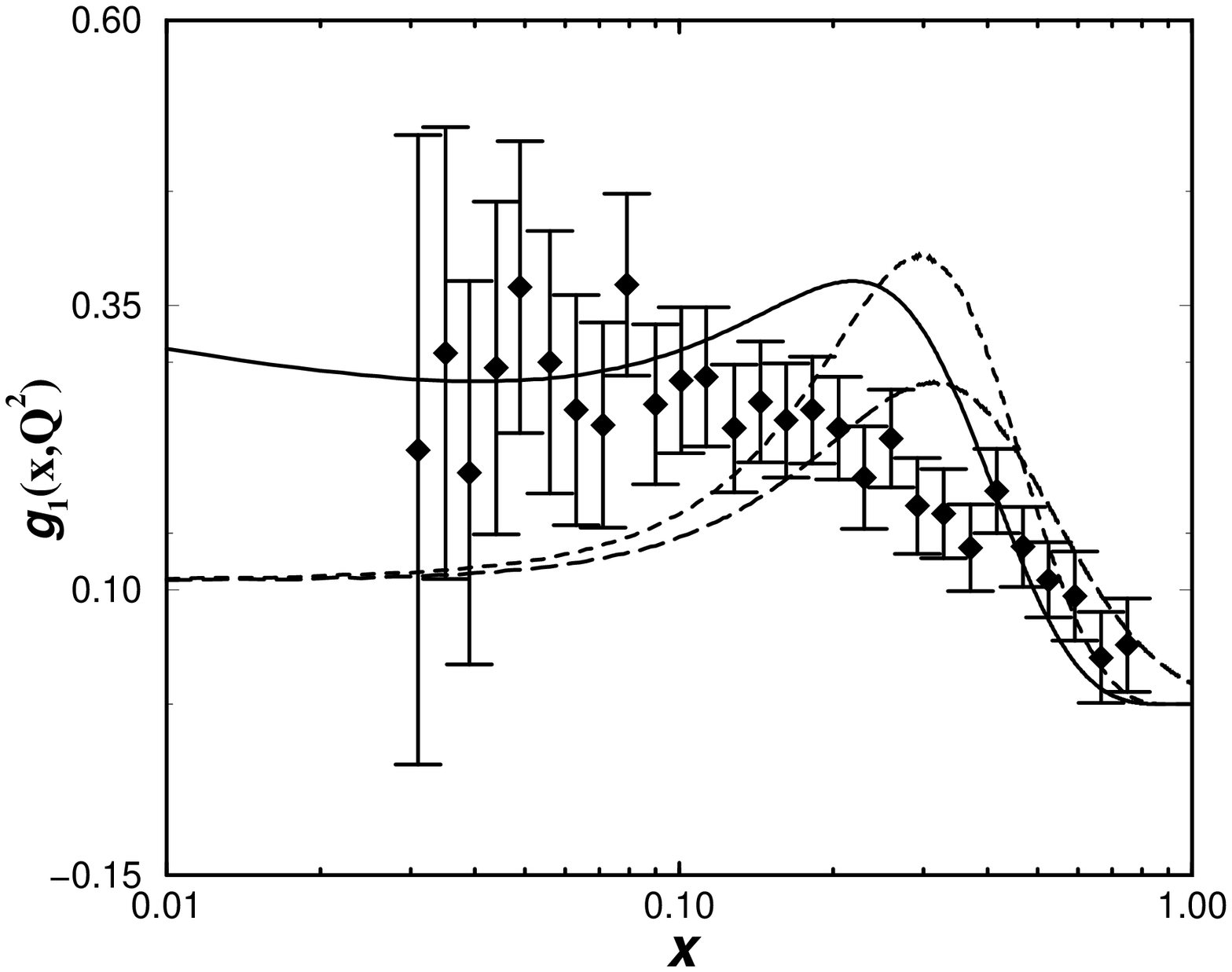,height=6.0cm,width=7.5cm}
\hspace{-1.0cm}
\epsfig{figure=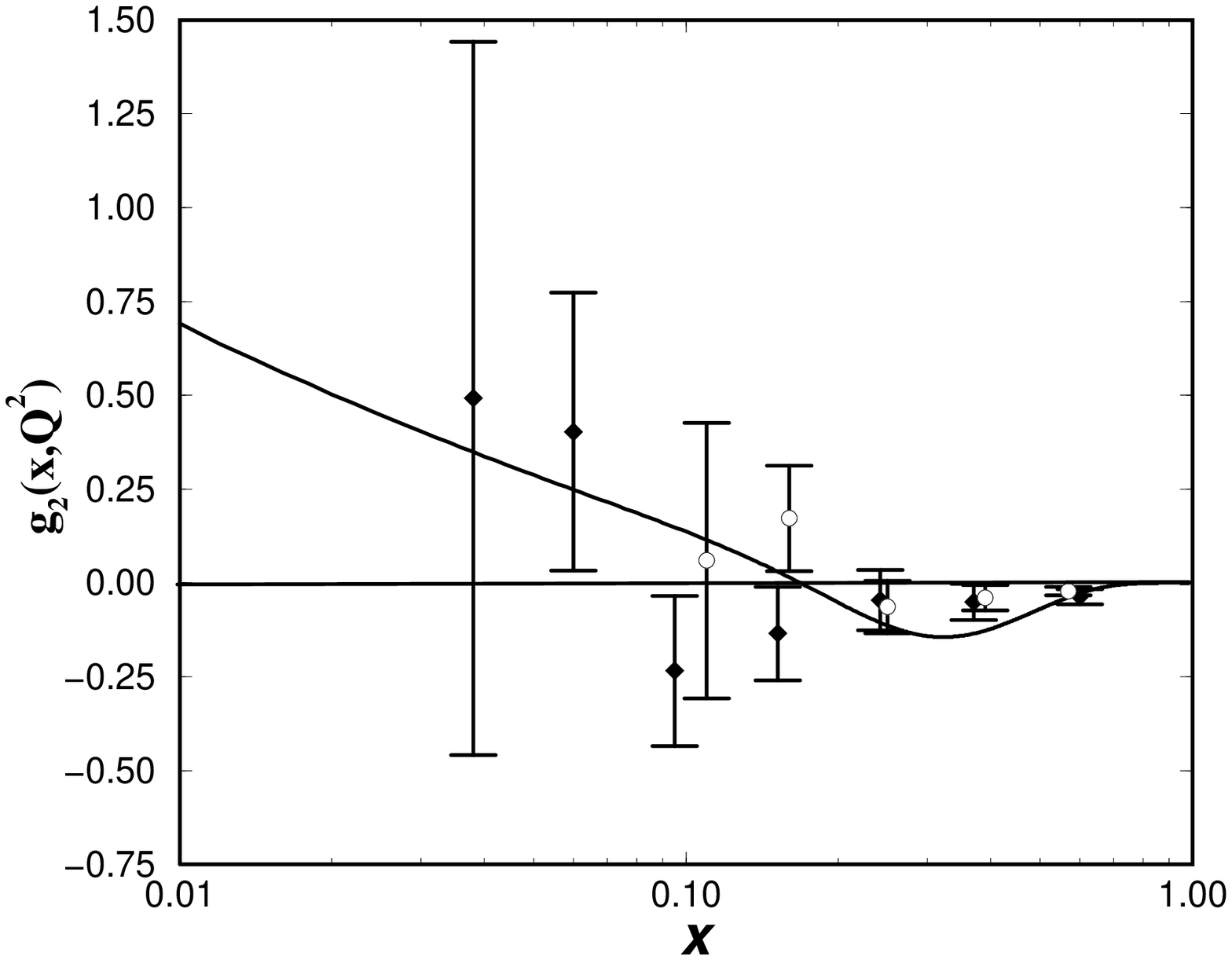,height=6.0cm,width=7.5cm}}
\caption{\label{fig_4}The polarized structure functions $g_1$
and $g_2$ after projection and QCD evolution. Left panel: 
The dashed (dotted) line denotes the (projected) low scale 
model prediction.}
\end{figure}

No low--scale approximation is available for the polarized structure 
function $g_2(x)$. We have therefore projected the predicted 
structure function onto the interval $x\in[0,1]$ \cite{Ja80} and 
subsequently performed a leading order QCD evolution to the scale 
of the experiment, see ref. \cite{We97b} for details. The resulting 
polarized structure functions are displayed in figure \ref{fig_4}.
Apparently the model reproduces the empirical data quite well, 
although the associated error bars are sizable.

\vspace{-0.2cm}
\section*{Conclusions}
\vspace{-0.2cm}

In this talk we have presented a calculation of nucleon 
structure functions within a chiral soliton model. We have 
argued that the soliton approach to the bosonized version of 
the NJL--model is most suitable since (formally) the required 
current operator is identical to the one in a free Dirac 
theory. Hence there is no need to approximate the current 
operator by {\it e.g.} performing a gradient expansion.
Although the calculation contains a few (well--motivated)
approximations it reproduces the gross features of the empirical
structure functions at low energy scales. This happens to be
the case for both the polarized as well as the unpolarized
structure functions.

Future projects will include to extend the valence quark 
approximation, improvements on the projection issue and the 
extension to flavor SU(3).

\end{document}